\documentclass[12pt]{article}
\usepackage{amsmath}
\usepackage{amssymb}
\usepackage{graphicx}

\newcommand{\lan}{\langle}
\newcommand{\ran}{\rangle}
\newcommand{\be}{\begin{equation}}
\newcommand{\bea}{\begin{eqnarray}}
\newcommand{\eea}{\end{eqnarray}}
\newcommand{\ee}{\end{equation}}
\newcommand{\vex}{{\bf x}}

\newcommand{{\tr}}{{\rm tr}}
\newcommand{{\const}}{{\rm const}}
\newcommand{{\adj}}{{\rm adj}}
\newcommand{{\PBC}}{{\rm PBC}}

\title{
\vspace{-10mm} \rightline{\small ITEP--PH--1--2003} \vspace{8mm}
Thermal gluo-magnetic vacuum of \\ $SU(N)$ gauge theory}
\author{
Nikita~O.~Agasian\thanks{e-mail: agasian@heron.itep.ru} \\
{\it Institute of Theoretical and Experimental Physics} \\
{\it 117218, Moscow, Russia}}
\date{}
\begin{document}

\maketitle
\vspace{1cm}
{\centerline {\bf Abstract}}
The magnetic sector of SU(N) Yang-Mills theory at finite temperature is studied.
At low temperatures, $T<2T_c$, the analytic  expressions
for the temperature dependence of the magnetic correlator, of the
magnetic gluon condensate and of the spatial string tension are obtained. Fair
agreement with lattice calculations for spatial string tension is obtained
for SU(2) and SU(3) gauge theories.
The relative contribution given by non-zero Matsubara modes to the spatial string
tension is calculated. At $T=2T_c$ this contribution is of the order of 5\%.
The behavior of magnetic correlator at high temperatures is investigated and it is
shown that gluo-magnetic condensate increases with temperature as
$\lan H^2 \ran (T) = \const \cdot g^8 (T) T^4$.

\vspace{1cm}


 \section{Introduction}

 It is known that at finite temperature non-Abelian SU(N)  gauge
 theories undergo  a deconfining phase transition passing from the
 low-energy glueball phase to the phase of hot gluon matter. At
 the critical point $T_c$ where the phase transition occurs the
 behavior of the thermodynamic properties of the system, such as
 energy density $\varepsilon$, specific heat, non-ideality
 $(\varepsilon -3P)/T^4$, etc., is drastically changed
 \cite{Engels:1994xj,Boyd:1996bx,Karsch:2001vs}.
 More than that, the phase transition in non-Abelian gauge
 theories is characterized by the radical rearrangement of the
 non-perturbative (NP) gluon vacuum.

 Recently the behavior of the gauge-invariant two-point
 correlation functions of the gauge fields across the
 deconfinement phase transition was investigated by numerical
 simulations on a lattice both for  the pure-gauge SU(3) theory
 and  for the full QCD with two flavors \cite{D'Elia:2002ck}. These data
 clearly demonstrate strong suppression of the electric component
 of the  correlator  above $T_c$ and persistence of the magnetic
 component. The magnetic correlator and its contribution to the
 gluon condensate is kept intact across the phase transition
 temperature,  while the   confining electric part abruptly
 disappears above $T_c$, so that the electric gluon condensate
 drops to zero at the deconfining phase transition point \cite{D'Elia:2002ck}.
 These results are completely in line with theoretical predictions
 of deconfining phase transition within the "evaporation model"
 \cite{Simonov:bc}  approach. Within the framework of the effective dilaton
 Lagrangian at finite temperature the discontinuity of the gluon
 condensate at $T=T_c$ was studied in \cite{Agasian:fn} (see also \cite{Sannino:2002wb}).

 At $T=T_c$ the physical string tension
 becomes zero. It was understood however that in non-Abelian gauge
 theories the Wilson loop for large space-like contours obeys the
 area law at arbitrary temperature. This phenomenon  is known as
 "magnetic confinement" and yields non-zero spatial string
 tension $\sigma_s(T) $ \cite{Borgs:qh}. The temperature dependence of
 $\sigma_s(T)$ was studied on a lattice for the pure-gauge SU(2)
 \cite{Bali:1993tz} and SU(3) \cite{Boyd:1996bx,Karsch:1994af} theories.
 It was shown there that
 $\sigma_s(T)$ smoothly passes through the phase transition
 temperature and increases with temperature. For temperatures
 larger than $2T_c$ the scaling behavior, $ \sqrt{\sigma_s(T)} /
 g^2T\sim $ const,  settles on \cite{Boyd:1996bx,Bali:1993tz,Karsch:1994af}.
This regime is imposed by the non-perturbative magnetic scale
$\sim g^2T$\cite{Linde:ts}. It is also firmly established that the 4d
SU(N) gauge theory at finite temperature is well described by an
effective 3d gauge theory with an adjoint Higgs field (the
so-called dimensional reduction) \cite{Ginsparg:1980ef,Kajantie:1997tt}. Moreover, the scalar
Higgs fields only weakly influence the physical properties of the
gauge theory magnetic sector
\cite{Bali:1993tz,Karsch:1994af,Kajantie:1997tt,Laine:1999hh,Hart:1999dj,Karsch:1998tx}.
Thus the behavior of
certain gluo-magnetic quantities in 3d are of direct relevance for
the behavior of these quantities in the 4d gauge theory at high
temperature. In particular, the  spatial string tension
$\sigma_s(T)$ at high  T coincides with good accuracy with lattice
results for the 3d string tension  $\sigma_3$ \cite{Karsch:1994af,Teper:gm}.

In the present paper we study the  magnetic sector of the 4d SU(N)
Yang-Mills theory at finite temperature. It is demonstrated that
thermal behavior of magnetic properties of the system is
qualitatively different in two temperature regions. We will define
$T<2T_c$ as low temperature region, and $T \ge 2T_c$ as high
temperature region\footnote{''2'' should not be considered as an
exact number. This matter is discussed in the last section.}. The
temperature behavior of the gluo-magnetic correlator, of the
magnetic  condensate and of   the spatial string tension in low
temperature region ($T<2 T_c$) are obtained  analytically. The
relative contribution of non-zero Matsubara modes to $\sigma_s(T)$ is calculated
and found to be of order of 5\% for pure $SU(3)$ gauge theory at
$T=2T_c$. The temperature interval corresponding to
the onset of the 3d regime of the Yang-Mills theory is investigated. We consider
the thermal character of the gluo-magnetic correlator in the
scaling region for the spatial string tension,
$\sqrt{\sigma_s}\sim g^2 T$. It is shown that the non-perturbative
gluo-magnetic condensate grows with temperature  as $\sim g^8(T)
T^4$ at high $T$.

\section{Gluo-magnetic correlator and spatial string tension}

Gauge-invariant non-local field correlators are of utmost
importance for our understanding of the NP QCD dynamics (see a
recent review \cite{Dosch:2000va} and references therein). The
gauge-invariant two-point correlators in the Yang-Mills vacuum are
defined as \cite{Dosch:1987sk,Dosch:ha,Simonov:1988mj}
\be
D_{\mu\nu\sigma\lambda} (x) = \lan g^2 \tr G_{\mu\nu} (x) \Phi
(x,0) G_{\sigma\lambda} (0) \Phi (0, x)\ran,
\label{1}
\ee
where
$G_{\mu\nu} =t^aG^a_{\mu\nu}$ is the field-strength tensor, $t^a$
are the SU(N) gauge group generators in the fundamental
representation, $\tr t^at^b=\frac12 \delta^{ab}$, and
\be
\Phi(x,y) = P\exp \{ ig \int^y_x dx_\mu A_\mu(x) \},~~ A_\mu = t^a
A^a_\mu
\label{2}
\ee
is the Schwinger parallel transporter with the
integration from $x$ to $y$ along a straight-line path. These
phase factors $\Phi$ are introduced to take care of the
gauge-invariance of the correlators. In general form bilocal
correlators (\ref{1}) are expressed in  terms of two independent
invariant functions of $x^2, D(x^2)$ and $D_1(x^2)$
\cite{Dosch:1987sk,Dosch:ha,Simonov:1988mj}.

At finite temperature the O(4) space-time symmetry is broken down
to the spatial O(3) symmetry and bilocal correlators are described
by independent electric and magnetic correlation functions.
On general grounds one can write for the magnetic
correlators the following expression $$ D^H_{ik} (x) = \lan g^2 \tr
H_i(x) \Phi (x,0) H_k(0) \Phi(0,x)\ran$$
\be
=\delta_{ik} ( D^H  + D^H_1) + (\delta_{ik}
-\frac{x_ix_k}{\vex^2})\vex^2 \frac{\partial
D_1^H}{\partial\vex^2},
\label{3}
\ee
where $H_i =\frac12
\varepsilon_{ijk} G_{jk}$ are the magnetic field operators. In a
well known way \cite{Dosch:ha} one can obtain the area law for large
spatial $W$-loops of size $L\gg\xi_m$, where $\xi_m$ is the magnetic
correlation length
\be
\lan W(C)\ran_{spatial} \sim \exp \{-\sigma_s
S_{min}\},
\label{4}
\ee
where the spatial string tension $\sigma_s$ is
\be
\sigma_s=\frac12 \int  d^2x \lan g^2 \tr H_n (x) \Phi(x,0) H_n(0)
\Phi(0,x)\ran
\label{5}
\ee
and $H_n$ is the component orthogonal to the space-like
plane of the contour.

Substituting (\ref{3}) into (\ref{5}) and keeping in mind that the terms with $D^H_1$
form the complete  derivative, one can present $\sigma_s$ in the
following form~\cite{Simonov:yn}
\be
\sigma_s = \frac12 \int d^2 x D^H(x).
\label{6}
\ee
Note that  only the term $D^H$ enters in the expression (\ref{6}) for $\sigma_s$.
At $T=0$ due to the $O(4)$ invariance electric and magnetic correlators coincide and
$\sigma_s=\sigma$, where $\sigma= (420 MeV)^2$ is a physical (temporal) string tension.

Note that magnetic correlator $D^H_{ik}$ can be expressed through matrix $U_{\adj}^{ab}$ in
the adjoint representation. Since integration in~(\ref{2}) goes along straight-line path,
connecting points $x$ and $y$, one has

\be
\begin{aligned}
D^H_{ik}(x,y) = \langle g^2 H_i^a(x) U_{\adj}^{ab}(x,y) H_k^b(y) \rangle, \\
U_{\adj}^{ab}(x,y) = 2 \tr \left( t^a \Phi(x,y) t^b \Phi(y,x) \right).
\end{aligned}
\ee

The functions $D^H$ and $D^H_1$ contain both perturbative $(\propto 1/x^4)$ and
nonperturbative ($\propto \exp \{-|x|/\xi_m\})$ terms and only non-perturbative parts of
the  correlators
contribute to the string tension (for the detailed discussion
of these questions see the review paper
\cite{Dosch:2000va}). The magnetic gluon condensate can  be also determined through the
NP contributions to the correlator and expressed in terms of the
functions $D^H$ and $D^H_1$ at $x^2=0$. The gluo-magnetic condensate can be written as
\be
D^H_{ii}(0) =\lan g^2 \tr H^2_i\ran \equiv \frac12 \lan H^2\ran
=3 (D^H(0)+D^H_1(0)),
\label{7}
\ee
where $\lan H^2\ran\equiv \lan (gH^a_i)^2\ran$.

The lattice data \cite{D'Elia:2002ck} demonstrate that at least in the
region of "measured" temperatures,
$T \lesssim 1.5 T_c$, the NP functions $D^H=B_0\exp \{-|x| /\xi_m\}$ and
$D^H_1=B_1\exp\{-|x|/\xi_m\}$ are almost independent of temperature. Correlation
length does not change and equals to it's value at $T=0$, and
$B_1\approx 0.05 B_0\ll B_0$. Only $B_0$ slightly grows with $T$, which leads to
the slow growth of condensate $\langle H^2 \rangle$ at low temperatures.

Let us see how one can obtain the analytical expressions for the
magnetic correlator and correspondingly
for the spatial string tension and magnetic gluon condensate
as functions of the temperature. It is well
known that the introduction of the temperature for the quantum field system in
thermodynamic equilibrium is equivalent to compactification along the euclidean
"time" component $x_4$ with the radius $\beta=1/T$ and imposing
the periodic boundary conditions (PBC)
for boson fields (anti-periodic for fermion ones).

Thermal vacuum averages are defined in a standard way

\begin{equation}
\langle \ldots \rangle_{\beta} = \frac{1}{Z_{\beta}}
\int_{\PBC} [D A] \ldots e^{-S_{\beta}[A]},
\label{eq_vac}
\end{equation}
where partition function is

\be
Z_{\beta}=\int_{\PBC} [D A] e^{-S_{\beta}[A]},
\quad S_{\beta}=\int_0^{\beta} dx_4 \int d^3 x L_{YM}.
\ee

As mentioned above, in the low temperature region
gluelump mass $M$ (inverse magnetic correlation
length $1/\xi_m$) does not depend on $T$. Thus we will use
zero-temperature expression for $D^H_{ik}(\vex,x_4)$ with PBC to obtain
thermal correlator
${\cal D}^H_{ik}=\langle g^2 H_i^a U_{\adj}^{ab} H_k^b \rangle_{\beta}$
built from gauge fields $H_i(\vex,x_4)$ and phase factor $U_{\adj}[A(\vex,x_4)]$ at a
particular "time" coordinate $x_4$

\be
{\cal D}^H_{ik}(\vex,\beta) =\sum^{+\infty}_{n=-\infty} D^H_{ik}(\vex, n\beta).
\label{8}
\ee

Usually, this approximation is valid in the case of free
noninteracting fields. For example, thermal Green function of free
scalar field is defined as sum over Matsubara modes of propagator
at $T=0$. At the same time, interaction changes particle mass and
makes it dependent on $T$. In our case in the region of low $T$ we
take into account only ''kinematic'' change of correlator, and
consider it as zero-temperature correlator with PBC. Physically
the expression (\ref{8}) is valid at $\beta>2\xi_m$, i.e. when the
two neighboring exponential functions do not overlap. We use the
value $1/\xi_m=1.5$ GeV and it is thus required that $T<1/2\xi_m
=750$ MeV. In the temperature region we consider in the present
section, $T<2T_c\sim 600$ MeV, this condition is fulfilled.
Moreover comparison of calculated below physical quantities with
lattice results confirm validity of expression~(\ref{8}) at
$T<2T_c$.

It should be noted that using Poisson's summation formula
\be
\frac{1}{\beta} \sum^{+\infty}_{n=-\infty} e^{i \omega_n x}
= \sum^{+\infty}_{n=-\infty} \delta(x-n\beta), \quad \omega_n = 2 \pi n/\beta
\ee
one finds, that~(\ref{8}) is a representation of magnetic correlator $D$ as a Fourier sum
over Matsubara frequencies

\be
\sum_n D^H_{ik}(\vex,n\beta)=\frac{1}{\beta} \sum_n \tilde D^H_{ik}(\vex,\omega_n),
\ee
where $\tilde D(\omega) = \int d\tau \exp(i\omega \tau) D(\tau)$
is a Fourier image of function $D$.

Thus consider the function

\be
f(\vex,\beta) =\sum^{+\infty}_{n=-\infty}
e^{-M\sqrt{\vex^2+n^2\beta^2}},~~ M\equiv 1/\xi_m.
\label{9}
\ee
In order to carry out the Matsubara summation over
the frequencies $\omega_n= 2\pi nT$ in (\ref{9}) use can be made of the
integral representation of the exponent in (\ref{9}).
This yields
\be
f(\vex,\beta) =\frac{1}{\sqrt{\pi}}\int^\infty_0 \frac{ds}{\sqrt{s}} e^{-s-M^2\vex^2/4s}
\sum^{+\infty}_{n=-\infty} e^{-M^2\beta^2 n^2/4s}.
\label{10}
\ee
The following summation equation holds
\be
\sum^{+\infty}_{n=-\infty}e^{-\frac{b^2n^2}{4s}}=
\frac{2\sqrt{\pi s}}{b} \sum^{+\infty}_{n=-\infty} e^{-\frac{4\pi^2n^2}{b^2}s}.
\label{11}
\ee
It enables to write (\ref{10}) in the form
\be
f(\vex,\beta) =\frac{2}{M\beta}\sum^{+\infty}_{n=-\infty}
\int^\infty_0 ds e^{-as-b/s},
\label{12}
\ee where
$a=1+4 \pi^2n^2/M^2\beta^2$ and $b=M^2\vex^2/4.$

The integral in the right-hand side of (\ref{12})  is expressed in
terms of the Macdonald function $K_1$. Finally we get
\be
f(\vex, T) =2 MT|\vex| \sum^{+\infty}_{n=-\infty} \frac{1}{\sqrt{M^2+\omega^2_n}}
K_1(|\vex|\sqrt{M^2+\omega^2_n}).
\label{13}
\ee
From the asymptotic behavior of function $f$ at short distances we can obtain the temperature
dependence of the magnetic gluon condensate, $\lan H^2\ran(T)= \lan H^2\ran f (|\vex | \to 0,T)$.
Using the behavior $K_1(x\to0) \to1/x$ one has
\be
\frac{\lan H^2\ran (T)}{\lan H^2\ran} = 2MT \sum^{+\infty}_{n=-\infty}\frac{1}{M^2+\omega^2_n}
= \coth \left (\frac{M}{2T}\right)
\label{14}
\ee
and at $T\ll M$
\be
\lan H^2\ran (T)/\lan H^2\ran = 1+2 e^{-M/T} +O(e^{-2M/T}).
\label{15}
\ee
This is completely in line with the lattice result \cite{D'Elia:2002ck}
showing the slow increase of the
magnetic gluon condensate at low $T\lesssim 1.5.T_c$.

Using the expression (\ref{6}) for  $\sigma_s$ and equation (\ref{13}) we get
\be
\frac{\sigma_s(T)}{\sigma} = \frac{4T}{M} \sum^{+\infty}_{n=-\infty}
\frac{1}{(1+\omega^2_n/M^2)^2}.
\label{16}
\ee
In arriving to (\ref{16}) use was made of the normalization
condition, $\sigma_s(0)=\sigma$, and $\int^{\infty}_0 x^2 dx
K_1(cx) =2/c^3$. Performing summation in (\ref{16}) we arrive to
the expression for the temperature dependence of the spatial
string tension at low temperatures

\be
\frac{\sigma_s(T)}{\sigma} =\frac{\sinh(M/T) +M/T}{\cosh (M/T) -1},
\label{17}
\ee
and $\sigma_s(T\ll M)/\sigma =1+2 (M/T+1) e^{-M/T}+O(e^{-2M/T})$.

From (\ref{16}) we can determine the relative contribution,
 $\Delta_n(T)\equiv \sigma_s^{n\neq 0} (T)/\sigma_s(T)$, of the
 non-zero Matsubara modes to the
spatial string tension as a function of $T$.
Sorting out from (\ref{16}) the term with $n=0$ we get

\be
\Delta_n (T) =1-\frac{4T}{M} \frac{\cosh(M/T)-1}{\sinh(M/T) + M/T}
\label{18}
\ee
It is clear that $\sigma^{n=0}_s(T)+\sigma_s^{n\neq0}(T) =
\sigma_s(T)$ and $\Delta_n (0)=1$.

Also one can obtain from~(\ref{16}) contribution of individual $n$-th non-zero Matsubara mode

\be
\frac{\sigma_s^n(T)}{\sigma_s^{n=0}(T)}
=\frac{M^4}{\omega_n^4}\frac{1}{\left(1+M^2/\omega_n^2\right)^2}
\ee
Thus, it can be seen that at $T>M/2 \pi$ contribution of $n$-th non-zero mode to the spatial
string tension as compared to the contribution of zero mode, $n=0$, very fast ''dies out'' with
increasing $T$

\be
\frac{\sigma_s^n(T)}{\sigma_s^{n=0}(T)}=\left(\frac{M}{2 \pi T}\right)^4\frac{1}{n^4} -
2 \left(\frac{M}{2 \pi T}\right)^6 \frac{1}{n^6} + \ldots
\ee

\section{High temperatures}

According to its definition the quantity $M=1/\xi_m$ is the
inverse magnetic correlation length. At $T=0$
the correlation lengths were determined by lattice calculations
\cite{Campostrini:1984xr}. On the other hand $M$
can be identified as the mass of the lowest magnetic gluelump with quantum
numbers $J^{PC}=1^{+-}$. Gluelumps \cite{Campbell:1985kp} are
not physical objects and their spectrum
cannot be measured in experiment. However, they play fundamental
role in the nonperturbative QCD
 since gluelump masses define the field correlators in the QCD vacuum and in
particular the string tension at $T=0$. The masses of gluelumps were
calculated analytically within the framework of the QCD sum rules
\cite{Dosch:1998th},
QCD string model \cite{Simonov:2000ky}
and computed numerically on the lattice \cite{Philipsen:2001ip}.
From all the calculations listed above one can
deduce the value of the correlation length in the interval $\xi\approx 0.1 \div 0.2$ fm.

In what follows we shall use the value $M=1.5$ GeV for inverse
magnetic correlation length at $T=0$. In Fig. 1 we present the dependence of the function
$\Delta_n(T)$ on $T/T_c$ ($T_c=270$ MeV for pure-gauge SU(3)  theory).
It follows that $\Delta_n(2 T_c)\simeq 0.05$ , i.e. the contribution non-zero
(non-static)  Matsubara modes
into $\sigma_s$ is about 5\%. This is of the same order as the
error of the lattice calculations
of the spatial string tension \cite{Boyd:1996bx,Bali:1993tz,Karsch:1994af}.
Thus at temperatures under consideration,
$T>2T_c$, the main contribution into the dynamics of the gauge theory magnetic  sector
comes from zero (static) Matsubara mode.

\begin{figure}[!ht]
\begin{picture}(350,240)
\put(20,15){\includegraphics{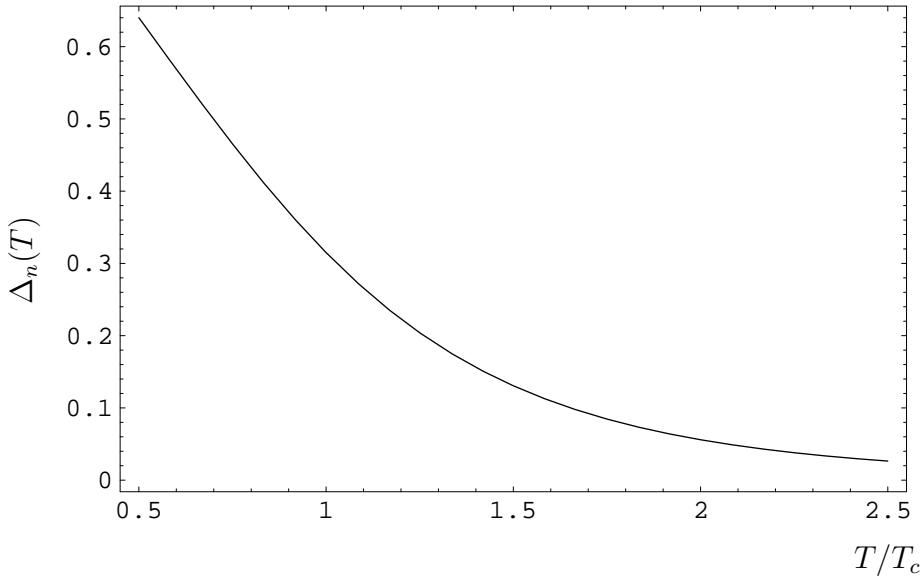}}
\put(0,100){\rotatebox{90}{$\Delta_n(T)$}}
\put(320,0){$T/T_c$}
\end{picture}
\caption{Relative contribution, $\sigma_s^{n\neq 0}/\sigma_s$, of the non-zero
Matsubara modes to the spatial string tension for pure-gauge SU(3) theory.}
\label{fig_1}
\end{figure}

Next we note  that at any $T$ the SU(N) gauge theory is in the phase of magnetic confinement.
Therefore at high $T$ the exponential form of the magnetic correlator remains unchanged.
 On the other hand the two-point
 correlation  function is determined by the amplitude and correlation length which are temperature
dependent.
Hence one can write the magnetic correlator at high temperature $T$  as
\be
{\cal D}^H_{ik} (\vex,T) =\frac16 \delta_{ik} \lan H^2\ran (T)
e^{-|\vex|/\xi_m(T)} +O\left(\vex^2\frac{\partial D^H_1}
{\partial\vex^2}\right).
\label{19}
\ee
The term $\sim O(...)$, as explained above, contributes neither
into gluo-magnetic condensate, nor into spatial string tension.
Then from (\ref{5}) and (\ref{19}) we get
\be
\sigma_s(T) =\frac{\pi}{6} \lan H^2\ran (T) \xi^2_m(T).
\label{20}
\ee

As it was shown in \cite{Boyd:1996bx,Bali:1993tz,Karsch:1994af}
the temperature dependence of the spatial
string tension $\sigma_s(T)$ at $T>2T_c$ with good accuracy coincides
with the behavior of the string tension
$\sigma_3$ in 3d Yang-Mills theory.  The three-dimensional
Yang-Mills theory is a superrenormalizable
theory and all physical quantities are determined by the only dimensionful
coupling constant  $g^2_3=g^2T$. It was shown in
\cite{Boyd:1996bx,Bali:1993tz,Karsch:1994af} that
\be
\sqrt{\sigma_s(T)}= c_\sigma g^2 (T) T,\label{21}
\ee
where use was made of the two-loop expression for $g^2(T)$
\be
\label{22}
\begin{aligned}
g^{-2}(T) =2 b_0 \ln \frac{T}{\Lambda_\sigma}+
\frac{b_1}{b_0}\ln \left(2\ln \frac{T}{\Lambda_\sigma}\right),\\
b_0=\frac{11N}{48\pi^2},~~~ b_1=\frac{34}{3} \left(\frac{N}{16\pi^2}\right)^2.
\end{aligned}
\ee
The two constants $c_\sigma$ and $\Lambda_\sigma$ were determined
using a two-parameter fit to lattice results. For the SU(2) gauge
theory $c_\sigma=0.369\pm 0.014,~~ \Lambda_\sigma = (0.076\pm
0.013)T_c$ \cite{Bali:1993tz}, while for SU(3) gauge theory
$c_\sigma=0.566\pm 0.013,~~ \Lambda_\sigma=(0.104\pm 0.009) T_c$
\cite{Boyd:1996bx}. In Figs. 2 and 3 we present the plots $\sigma_s
(T)/\sigma$ for SU(2) (Fig.2) and SU(3) (Fig.3) gauge theories as
functions of $T/T_c~~ (T_c^{SU(2)}=290$ MeV, $T_c^{SU(3)}=270$
MeV). Solid lines correspond to Eq.(\ref{17}) and the dashed ones to
Eqs.(\ref{21},\ref{22}). The lattice data are from Refs.\cite{Boyd:1996bx,Bali:1993tz}.

Further, from (\ref{20}) it is possible to find $\lan H^2\ran
(T)$. The inverse magnetic correlation length $1/\xi_m$ at high
temperature behaves as
\be
1/\xi_m(T) = c_m g^2(T) T,
\label{23}
\ee
where $c_m$ is some constant.
Using (\ref{20}), (\ref{21}) and (\ref{23}) we obtain the following expression for the
temperature dependence of the gluo-magnetic condensate
\be
\lan H^2\ran (T) =c_H g^8 (T) T^4,
\label{24}
\ee
and
\be c_H=\frac{6}{\pi} c^2_\sigma c^2_m.
\label{25}
\ee
From~(\ref{25}) one can estimate $c_H$. For instance in $SU(2)$ Yang-Mills theory,
$c_{\sigma}\approx 0.37$\cite{Bali:1993tz}, $c_m\approx 0.92$
\cite{Petreczky:1999ab}, and one finds $c_H \approx 0.22$.
However, the value of $c_H$ should be found from numerical
 simulations on a lattice for thermal magnetic correlator.

The result (\ref{24}) seems quite natural from the 3d gauge theory
standpoint. As was already indicated, the thermal behavior of the
physical quantities in the magnetic sector of 4d SU(N) gauge theory at high $T$ coincides
 with the behavior of corresponding quantities in the 3d gauge theory.
 Then the NP gluo-magnetic condensate
is determined by the value of the dimensionful coupling constant $g_3$
and $\lan H^2\ran (T) =$ const $g^8_3$.\footnote{Strictly speaking, in
superrenormalizable 3d Yang-Mills theory vacuum averages either are equal to zero,
or are defined through corresponding power of coupling constant $g_3$. Disappearance of
magnetic condensate at high temperatures means that $\sigma_s=0$, and therefore
contradicts lattice data on magnetic confinement.}
Therefore at high temperatures (in the scaling region for
the spatial string tension) the gluo-magnetic correlator has the form
\be
{\cal D}^H_{ik} (\vex, T) =\frac16 \delta_{ik} c_H g^8 (T) T^4 e^{-c_m
g^2(T) T |\vex|}.
\label{26}
\ee
The amplitude of this correlator rises as $T^4$ while the correlation length drops with
growing temperature as $1/T$ in the high temperature region.

\begin{figure}[!ht]
\begin{picture}(440,185)
\put(20,35){\includegraphics[width=210pt,height=140pt]{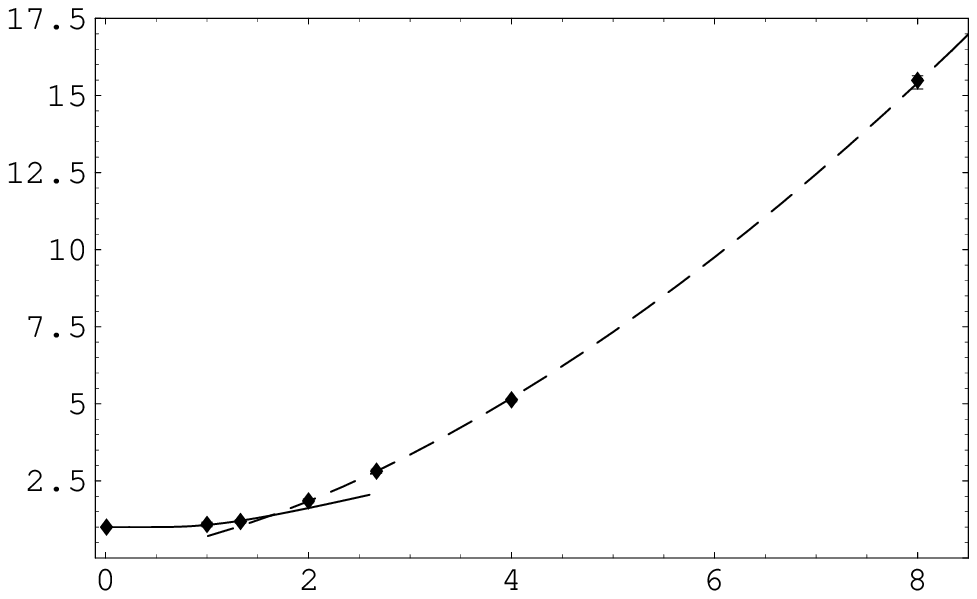}}
\put(240,35){\includegraphics[width=210pt,height=140pt]{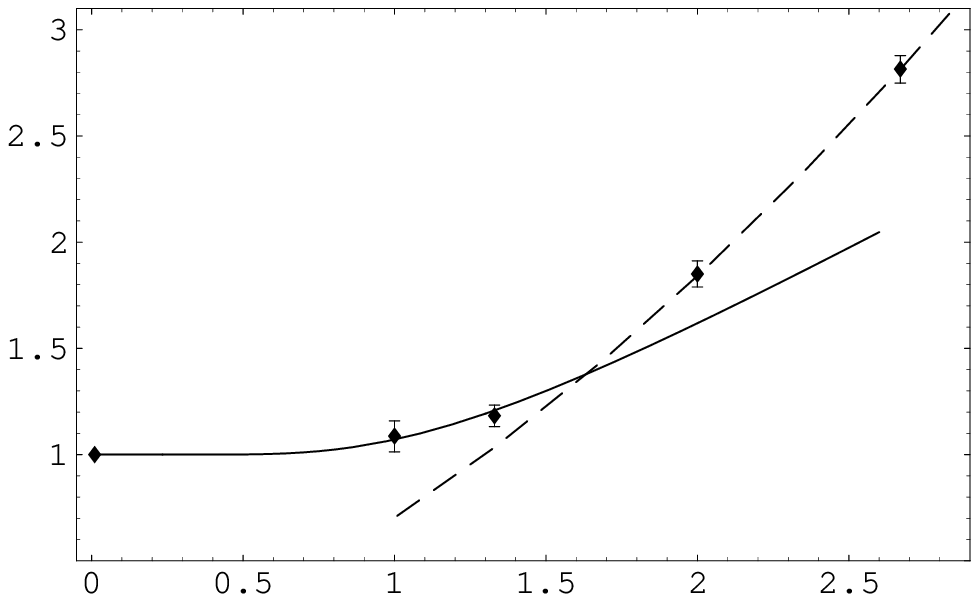}}
\put(0,90){\rotatebox{90}{$\sigma_s(T)/\sigma$}}
\put(210,20){$T/T_c$}
\put(430,20){$T/T_c$}
\end{picture}
\caption{Spatial string tension $\sigma_s (T)/\sigma$ for SU(2) gauge theory as
function of $T/T_c$. Solid lines correspond to Eq.(\ref{17}) and the dashed ones to
Eqs.(\ref{21},\ref{22}). The lattice data are from Refs.\cite{Bali:1993tz}.}
\label{fig_2}
\end{figure}

\begin{figure}[!ht]
\begin{picture}(440,185)
\put(20,35){\includegraphics[width=210pt,height=140pt]{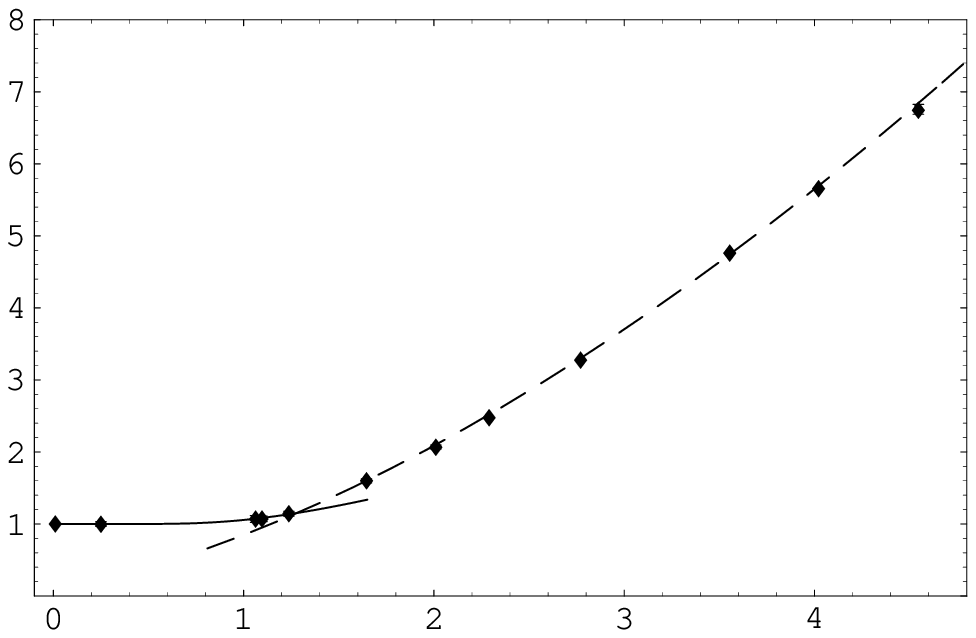}}
\put(240,35){\includegraphics[width=210pt,height=140pt]{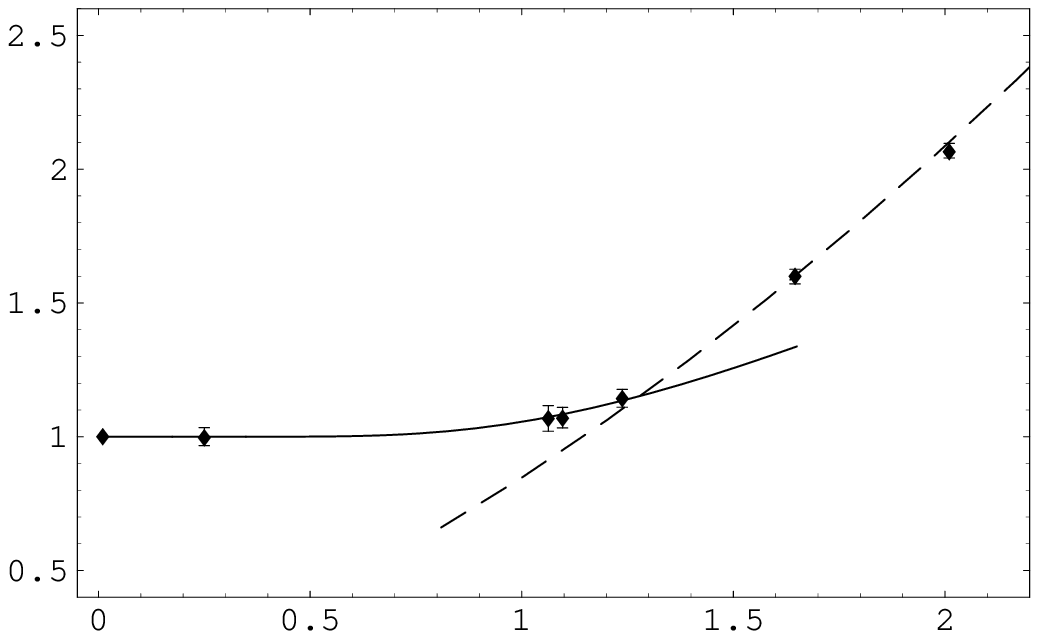}}
\put(0,90){\rotatebox{90}{$\sigma_s(T)/\sigma$}}
\put(200,20){$T/T_c$}
\put(430,20){$T/T_c$}
\end{picture}
\caption{Spatial string tension $\sigma_s (T)/\sigma$ for SU(3) gauge theory as
function of $T/T_c$. Solid lines correspond to Eq.(\ref{17}) and the dashed ones to
Eqs.(\ref{21},\ref{22}). The lattice data are from Refs.\cite{Boyd:1996bx}.}
\label{fig_3}
\end{figure}

\section{Discussions and conclusion}

The results which were obtained above allow to estimate the temperature region
 within which the change of the regime in the magnetic sector of SU(N) gauge theory occurs and
 it becomes possible to describe the dynamics in terms of the  static correlation
 functions, i.e. in terms of the 3d gauge theory. On physical grounds it is clear that
when the compactification radius $\beta$ (the inverse temperature
$1/T$) becomes much smaller than the typical dimension $l$ of the
system (which in turn is determined either by magnetic string
thickness, or by the radius of the magnetic gluelump)
$l=1/\sqrt{\sigma_s (T)}$, the correlators "cease to feel the time
coordinate $x_4$" and become pure static.

Thus the condition $T=\sqrt{\sigma_s(T)}$ enables to estimate the temperature at which the
change of the regimes occurs. Consider for definiteness  the SU(3) gauge theory. If one starts
from the low temperature region
then from the numerical solution of the equation
$T_-=\sqrt{\sigma_s(T_-)}$ and from Eq. (\ref{17})
for $\sigma_s (T)$ one finds $T_-=1.42 T_c$.
Alternatively if one starts from the high temperature
region then the solution of the equation $T_+=\sqrt{\sigma_s(T_+)}$ with
$\sigma_s(T)$ defined by (\ref{21},\ref{22}) gives $T_+=2.74 T_c$.
 Thus we may conclude that within the interval $T_-
<T<T_+$ the change of the regimes takes place and
transition to the reduced 3d Yang-Mills theory occurs.
Therefore the commonly accepted value $2T_c$ seems quite
natural from the outlined physical point of view.

The growth of the magnetic condensate, $\lan H^2 \ran (T) \propto
T^4$, with the increase of the temperature is thermodynamically
advantageous for the quantum field system. The vacuum energy
density is connected via the trace anomaly\footnote{Connection between trace anomaly
and thermodynamic pressure is considered in
Refs.~\cite{Leutwyler:cd,Ellis:1998kj,Drummond:1999si,Agasian:2001bj}}
with the magnetic
condensate by the relation $\varepsilon_{vac} =\langle \theta_{00} \rangle
=-(b/64\pi^2)\lan H^2\ran,~~b=11N/3,$ and the increase of $\lan
H^2\ran $ lowers the vacuum energy density.

In the present paper we have investigated  the magnetic sector of
SU(N) Yang-Mills theory at finite temperature. At low
temperatures, $T<2T_c$, the analytic  expressions were obtained
for the temperature dependence of the magnetic correlator,
magnetic gluon condensate and the spatial string tension. Fair
agreement with lattice calculations for $\sigma_s(T)$ was obtained
for SU(2) and SU(3) gauge theories.
It is demonstrated, that the contribution of $n$-th non-zero
Matsubara mode to the spatial
string tension as compared to the contribution of zero mode,
$n=0$, very fast ''dies out'' with
increasing temperature $\sim (M/2\pi n T)^4$.
The relative contribution
given by non-zero Matsubara frequencies to the spatial string
tension was calculated. At $T=2T_c$ this contribution is about
5\%. The estimate was
presented of the temperature region corresponding to the transition to the
description of the nonperturbative dynamics in the magnetic
sector in terms of the $3d$ correlation functions.
We have found
the behavior of magnetic correlator at high $T>2T_c$ and have
shown that gluo-magnetic condensate increases with temperature as
$\lan H^2 \ran (T) = c_H g^8 (T) T^4$.

\begin{center}
{\bf ACKNOWLEDGMENTS} \\
\end{center}

I am grateful to Yu.A.~Simonov for discussions and useful comments, and
I thank B.L.~Ioffe and V.A.~Rubakov for discussions of results.
The financial support INTAS grant N 110 is gratefully
acknowledged.


\begin{thebibliography}{99}

\bibitem{Engels:1994xj}
J.~Engels, F.~Karsch and K.~Redlich,
Nucl.\ Phys.\ B {\bf 435} (1995) 295 [hep-lat/9408009].

\bibitem{Boyd:1996bx}
G.~Boyd, J.~Engels, F.~Karsch, E.~Laermann, C.~Legeland,
M.~Lutgemeier and B.~Petersson,
Nucl.\ Phys.\ B {\bf 469} (1996) 419 [hep-lat/9602007].

\bibitem{Karsch:2001vs}
F.~Karsch,
Nucl.\ Phys.\ A {\bf 698} (2002) 199 [hep-ph/0103314];
%
Lect.\ Notes Phys.\  {\bf 583} (2002) 209 [hep-lat/0106019];
%
E.~Laermann and O.~Philipsen, ``Status of lattice QCD at finite
temperature,'' hep-ph/0303042.

\bibitem{D'Elia:2002ck}
M.~D'Elia, A.~Di Giacomo and E.~Meggiolaro, ``Gauge-invariant
field-strength correlators in pure Yang-Mills and full  QCD at
finite temperature,'' hep-lat/0205018.

\bibitem{Simonov:bc}
Y.~A.~Simonov,
JETP Lett.\  {\bf 55} (1992) 627;
%
Phys.\ Atom.\ Nucl.\  {\bf 58} (1995) 309 [hep-ph/9311216].

\bibitem{Agasian:fn}
N.~O.~Agasian,
JETP Lett.\  {\bf 57} (1993) 208;
%
N.~O.~Agasian, D.~Ebert and E.~M.~Ilgenfritz,
Nucl.\ Phys.\ A {\bf 637} (1998) 135 [hep-ph/9712344].

\bibitem{Sannino:2002wb}
F.~Sannino,
Phys.\ Rev.\ D {\bf 66} (2002) 034013 [hep-ph/0204174].

\bibitem{Borgs:qh}
C.~Borgs,
Nucl.\ Phys.\ B {\bf 261} (1985) 455;
%
E.~Manousakis and J.~Polonyi,
Phys.\ Rev.\ Lett.\  {\bf 58} (1987) 847.

\bibitem{Bali:1993tz}
G.~S.~Bali, J.~Fingberg, U.~M.~Heller, F.~Karsch and K.~Schilling,
Phys.\ Rev.\ Lett.\  {\bf 71} (1993) 3059 [hep-lat/9306024].

\bibitem{Karsch:1994af}
F.~Karsch, E.~Laermann and M.~Lutgemeier,
Phys.\ Lett.\ B {\bf 346} (1995) 94 [hep-lat/9411020].

\bibitem{Linde:ts}
A.~D.~Linde,
Phys.\ Lett.\ B {\bf 96} (1980) 289;
%
D.~J.~Gross, R.~D.~Pisarski and L.~G.~Yaffe,
Rev.\ Mod.\ Phys.\  {\bf 53} (1981) 43.

\bibitem{Ginsparg:1980ef}
P.~Ginsparg,
Nucl.\ Phys.\ B {\bf 170} (1980) 388;
%
T.~Appelquist and R.~D.~Pisarski,
Phys.\ Rev.\ D {\bf 23} (1981) 2305;
%
S.~Nadkarni,
Phys.\ Rev.\ D {\bf 27} (1983) 917;
%
N.~P.~Landsman,
Nucl.\ Phys.\ B {\bf 322} (1989) 498;
%
K.~Kajantie, M.~Laine, K.~Rummukainen and M.~E.~Shaposhnikov,
Nucl.\ Phys.\ B {\bf 458} (1996) 90 [hep-ph/9508379].

\bibitem{Kajantie:1997tt}
K.~Kajantie, M.~Laine, K.~Rummukainen and M.~E.~Shaposhnikov,
Nucl.\ Phys.\ B {\bf 503} (1997) 357 [hep-ph/9704416].

\bibitem{Laine:1999hh}
M.~Laine and O.~Philipsen,
Phys.\ Lett.\ B {\bf 459} (1999) 259 [hep-lat/9905004].

\bibitem{Hart:1999dj}
A.~Hart and O.~Philipsen,
Nucl.\ Phys.\ B {\bf 572} (2000) 243 [hep-lat/9908041].

\bibitem{Karsch:1998tx}
F.~Karsch, M.~Oevers and P.~Petreczky,
Phys.\ Lett.\ B {\bf 442} (1998) 291 [hep-lat/9807035];
%
A.~Cucchieri, F.~Karsch and P.~Petreczky,
Phys.\ Lett.\ B {\bf 497} (2001) 80 [hep-lat/0004027].

\bibitem{Teper:gm}
M.~Teper,
Phys.\ Lett.\ B {\bf 311} (1993) 223.

\bibitem{Dosch:2000va}
A.~Di Giacomo, H.~G.~Dosch, V.~I.~Shevchenko and Y.~A.~Simonov,
Phys.\ Rept.\  {\bf 372} (2002) 319 [hep-ph/0007223].

\bibitem{Dosch:1987sk}
H.~G.~Dosch,
Phys.\ Lett.\ B {\bf 190} (1987) 177.

\bibitem{Dosch:ha}
H.~G.~Dosch and Y.~A.~Simonov,
Phys.\ Lett.\ B {\bf 205} (1988) 339.

\bibitem{Simonov:1988mj}
Y.~A.~Simonov,
Nucl.\ Phys.\ B {\bf 324} (1989) 67.

\bibitem{Simonov:yn}
Y.~A.~Simonov,
JETP Lett.\  {\bf 54} (1991) 249;
%
``Hot nonperturbative QCD,'' Lectures given at International
School of Physics, 'Enrico Fermi',
Varenna, Italy, 1995. In *Varenna 1995, Selected topics in
nonperturbative QCD* 319-337 [hep-ph/9509404].

\bibitem{Campostrini:1984xr}
M.~Campostrini, A.~Di Giacomo and G.~Mussardo,
Z.\ Phys.\ C {\bf 25} (1984) 173;
%
A.~Di Giacomo and H.~Panagopoulos,
Phys.\ Lett.\ B {\bf 285} (1992) 133;
%
A.~Di Giacomo, E.~Meggiolaro and H.~Panagopoulos,
Nucl.\ Phys.\ B {\bf 483} (1997) 371 [hep-lat/9603018];
%
G.~S.~Bali, N.~Brambilla and A.~Vairo,
Phys.\ Lett.\ B {\bf 421} (1998) 265 [hep-lat/9709079].

\bibitem{Campbell:1985kp}
N.~A.~Campbell, I.~H.~Jorysz and C.~Michael,
Phys.\ Lett.\ B {\bf 167} (1986) 91;
%
I.~H.~Jorysz and C.~Michael,
Nucl.\ Phys.\ B {\bf 302} (1988) 448.

\bibitem{Dosch:1998th}
H.~G.~Dosch, M.~Eidemuller and M.~Jamin,
Phys.\ Lett.\ B {\bf 452} (1999) 379 [hep-ph/9812417];
%
M.~Eidemuller, H.~G.~Dosch and M.~Jamin,
Nucl.\ Phys.\ Proc.\ Suppl.\  {\bf 86} (2000) 421
[hep-ph/9908318].

\bibitem{Simonov:2000ky}
Y.~A.~Simonov,
Nucl.\ Phys.\ B {\bf 592} (2001) 350 [hep-ph/0003114].

\bibitem{Philipsen:2001ip}
O.~Philipsen,
Nucl.\ Phys.\ B {\bf 628} (2002) 167 [hep-lat/0112047];
%
M.~Laine and O.~Philipsen,
Nucl.\ Phys.\ B {\bf 523} (1998) 267 [hep-lat/9711022].

\bibitem{Petreczky:1999ab}
P.~Petreczky, ``Screening in hot non-Abelian plasma,''
hep-ph/9907247.


\bibitem{Leutwyler:cd}
H.~Leutwyler, ``Deconfinement And Chiral Symmetry,''
{\it Prepared for Workshop on QCD: 20 Years Later, Aachen,
Germany, 9-13 Jun 1992}.

\bibitem{Ellis:1998kj}
P.~J.~Ellis, J.~I.~Kapusta and H.~B.~Tang,
Phys.\ Lett.\ B {\bf 443} (1998) 63 [nucl-th/9807071].

\bibitem{Drummond:1999si}
I.~T.~Drummond, R.~R.~Horgan, P.~V.~Landshoff and A.~Rebhan,
Phys.\ Lett.\ B {\bf 460} (1999) 197 [hep-th/9905207].

\bibitem{Agasian:2001bj}
N.~O.~Agasian,
Phys.\ Lett.\ B {\bf 519} (2001) 71 [hep-ph/0104014];
%
JETP Lett.\  {\bf 74} (2001) 353 [hep-ph/0104193];
%
hep-ph/0212392.

\end{thebibliography}
\end{document}